\documentstyle[12pt]{article}
\begin{document}

\begin{center}

\bf{TIME VARIATIONS OF SOLAR NEUTRINO. 
THE MAIN ARGUMENTS PRO AND SOME INFERENCES}
\end{center}
\smallskip
\centerline{Rivin Yu.R., Obridko V.N.}
\smallskip
\begin{center}

\bf{I.  Arguments in support of time variations of the high-speed solar neutrino 
flux, recorded with the Homestake detector.}
\end{center}
\smallskip

1. In a separate run of measurements of solar neutrino by the Homestake detector, 
the ratio of the number of $^{37}Ar$ atoms per day (signal) to the average measuring 
error (noise) is $\approx1$. For the annual mean values, the signal--to-noise ratio is 
$\approx3-4$, which allows us to analyze cyclic (and, less reliably, ``quasi-biennial'') 
variations of the solar neutrino flux, Pn (Rivin et al., 1983; Rivin, 1989-1993; Rivin \&  
Obridko, 1997).

2. According to interpretation of the $P_{\nu}$ curve and its spectrum, suggested 
in the mentioned works, these variations have two independent sources. One is the 
processes inside the convection zone that produce oscillations with a quasi-period 
$T\approx1$ years. Another source is associated with thermal processes in the core that 
have a characteristic period of $\approx2-5$ years, i.e. ``quasi-biennial'' variations 
(Sacurai, 1979; Sacurai \& Hasegawa, 1985; Rivin et al., 1983). 

	Up to 1985, the neutrino flux had displayed a good correlation with the Wolf 
numbers, W, changing in anti-phase with a period of 11-years (Rivin et al., 1983; Rivin, 
1993; Davis, 1984). After 1985 (in solar cycle 22), the correlation broke: the maximum 
values of $W$ in cycles 21 and 22 were approximately equal, whereas the maximum of 
$P_{\nu}$  in cycle 22 was significantly lower, than in the previous one. The correlation 
could be somewhat improved by smoothing the original curves (removing the short-
periodic oscillations), but it still remained poor after 1985 non only in the Homestake, but 
also in the Kamiokande-II data. This fact raised serious doubts as to whether any time 
variations of $P_{\nu}$ did exist at all. However the reason for such discrepancy was 
explained later in (Obridko \& Rivin, 1995, 1996; Rivin \& Obridko, 1997; Rivin, 1997). 
It was shown that the Wolf numbers did not adequately represent cyclic variations of the 
modulus of the deep quadrupole solar magnetic field, $\mid B_q\mid$, that proved to 
conserve correlation with $P_{\nu}$. The correlation even became better during the past 
20 years. The fact is that the amplitude of $\mid B_q\mid$ variations, modulating the 
neutrino flux, is by several fold smaller in cycle 22, than in cycle 21. This ratio of 
amplitudes in two successive cycles differs much from that of the Wolf number series. 

	Later on, this work was used to construct a model of 11-year variations of $\mid 
B\mid$ (Obridko \& Rivin, 1996c; Rivin, 1997a, 1998a-c). The model involves three 
spatial coordinates in addition to the temporal one, and provides a simplified mechanism 
that describes observations better, that the available theoretical models. 

	The observation series of solar neutrino is too short to justify any definite 
conclusions, however the coincidence in shape of the $P_{\nu}$ and $\mid B_q\mid$ 
amplitude variations provides an additional evidence in support of our hypothesis of the 
11-year cycle and other variations of $P_{\nu}$.

	Note that the present-day argument pro and contra time variations of $P_{\nu}$ 
has historic analogy. G.Schwabe, who discovered the 11-year cycle of solar activity in 
1843, could not publish his results for several years, because authoritative scientists of 
that time believed them implausible, taking into account that the observed and calculated 
diurnal Wolf numbers displayed a scatter far beyond the annual mean values. The 
publication was finally promoted by A.Humboldt, however another 50--60 years had to 
pass before the solar cycle was broadly recognized by the scientific community. We hope 
that recognition of the 11-year cycle and other variations of $P_{\nu}$ will not take long.

	The quasi-biennial variation of $P_{\nu}$ is much more obscured by the 
recording errors, than the 11-year cycle, and besides, it is additionally distorted due to a 
break in operation of the detector in 1985--1986. Nevertheless Rivin (1993) could reveal 
a good correlation of $P_{\nu}$ with the respective variations of Wolf numbers and a 
somewhat worse correlation with $\mid B_q\mid$. 

	Thus, the correlation of $P_{\nu}$ with  the Wolf numbers and $\mid B_q\mid$ 
over two periods ($\sim11$ years and $\sim2-5$ years) was established reliably enough 
to make inferences of real time variations of the neutrino flux and even of its modulation 
sources. 

	Another important feature of time variations of the neutrino flux is described in 
(Rivin et al., 1983; Rivin, 1989, 1993). The amplitudes of $P_{\nu}$ variations with 
$T\approx11$ years and $T\approx2-5$ years are approximately equal, unlike the case of 
$W$ and $\mid B_q\mid$, where the amplitude of the 11--year cycle is much larger, than 
the amplitude of oscillations with a shorter period. This fundamental difference was not 
taken into account in statistical studies (e.g. in the work by Bahcall (1989), in the chapter 
devoted to neutrino time variations). This erroneously resulted in the conclusion that time 
variations in $P_{\nu}$ were absent. The analysis would have been more correct if short-
period variations had been filtered out from the data series before processing. Besides, 
taking into account all said above, it would be reasonable to use as a characteristic of the 
solar magnetic field (especially after 1986) the modulus of the large-scale field, measured 
directly in the photosphere, rather than the Wolf numbers (Obridko \& Rivin, 1995a, 
1996a).

	3. The works by Akhmedov (1997) and Masseti \& Storini (1996) served a 
starting point for our analysis of seasonal variations in $P_{\nu}$. As a result, an annual 
wave with the extrema at equinoxes (in spring higher and in autumn lower than at the 
solstices) was isolated by two different methods. It was suggested to be due to the 
Voloshin-Vysotsky-Okun mechanism (VVO) (Voloshin et al., 1986), working in the 
magnetic field with a significant spatial asymmetry (Rivin \& Obridko, 1998). This 
variation, however, was not revealed in the data from Kamiokande-II and 
SuperKamiokande quoted by Suzuki (1996). The explanation may be as follows: as a 
result of modulation by the 11--year cycle the amplitude of the annual wave is much 
smaller in the years of minimum, than in the years of maximum of cycle 21, and it may 
become pronounced again in a year or two. 

Thus, the joint analysis of the Homestake Pn data and the $\mid B_q\mid$ 
measurements in the photosphere allows us to make a conclusion of two probable 
mechanisms in the Sun, responsible for modulation of the neutrino flux, $P_{\nu}$. One 
is the modulus of the magnetic quadrupole, generated at the base of the transition layer 
between the convection and the radiative zones, where the differential rotation is changed 
by the rigid one (here, the 11--year cycle and the annual wave take their origin). Another 
source is associated with the thermal fluctuations of the core (quasi--biennial variations).

	Since the quadrupole magnetic field is likely to rotate mainly with a period 
$T\approx27$ days, one should expect to observe the same period in $P_{\nu}$. This 
inference can probably be verified in future using the SuperKamiokande and other up-to-
date detectors. Preliminary diurnal $P_{\nu}$ data from the SuperKamiokande detector 
were given by Suzuki (1996) for 175 days after 1.04.1996. These data did not reveal any 
pronounced fluctuations with $T\approx27$ days, though the scatter of intividual values 
was quite significant. However it may be due to that fact that, 1996 being the year of 
solar minimum, the amplitude of the 27-day variation of $\mid B_q\mid$ (and hence he 
corresponding $P_{\nu}$ wave) was too small to be positively detected. In a year or two 
the situation may change.

	Unfortunately, the data from other detectors (e.g. Neutrino--96) are practically 
useless for the analysis of $P_{\nu}$ and are not easily accessible. 

\medskip
\begin{center}
\bf{II. Some speculations on the $P_{\nu} (t)$ modulation by $\mid B_q\mid$}
\end{center}
\smallskip

	The transverse magnetic field, modulating the neutrino, flux is usually estimated 
from the sunspot magnetic field, measured in the photosphere (Voloshin et al., 1986). 
However this procedure does not take into account the following:

	1) The nature of sunspot magnetic fields is not quite clear. Are they ``fragments'' 
of the toroidal field that tear off and emerge as magnetic tubes in the photosphere, or do 
they proceed from the cut-off fields in the magnetic tube generation region? In the latter 
case, their intensity must be much lower, than the intensity of the original toroidal field. 
For more than a hundred year history of observations, the sunspot magnetic fields did not 
display cyclic variations of their intensity (H) that are characteristic of the corresponding 
toroidal dynamo field ($B$). Some authors (e.g., Vitinsky et al., 1986) argue that such 
periodicity in $H$ is absent, and the cyclic variation of solar activity is only manifested 
in the changing number of sunspots. This means that the sunspot magnetic fields are most 
likely due to the cut-off fields in the magnetic flux generation region (the generation 
mechanism was described, in particular, by Parker (1975)). In this case, the obtained 
estimate seems to be merely the lower limit of the real magnetic field value near the 
generation region.

	2) Does the intensity of sunspot fields correspond to that of the fields at the base 
of the convection zone? This question arises in connection with the new model of cyclic 
variations of $\mid B_q\mid$, suggested by Obridko \& Rivin (1996c) and developed by 
Rivin (1997, 1998).

	The model suggests the existence of two independent, but related regions of 
magnetic field generation in the Sun. The upper dynamo mechanism at the center of the 
convection zone generates the field of quasi-axial dipole, changing with a period 
$T\approx22$ years. This is where $\sim99\%$ of the sunspots observed in the 
photosphere occur. Taking into account all said above about the possible origin of 
sunspot fields, one can suggest that the intensity of the toroidal field, generated by the 
aw-dynamo, may reach in this region $\ge1-5$ kG (Steshenko, 1967). The lower dynamo 
of the quadrupole magnetic field with a probable period $T\approx27$ days operates at 
the bottom of the convection zone, where the rotation is rigid. (It is $\alpha\alpha$ or 
some other dynamo mechanism, but not $\alpha\omega$). Due to a large radial density 
gradient in the convection zone (five to six orders of magnitude), part of the upper field is 
``pumped'' away from the generation region to the base of the convection zone as a result 
of anisotropic turbulent pumping (ATP). As the ATP mechanism is nonlinear, the 
pumping is accompanied by a detection (a cycle with $T\approx11$ years appears) and 
enhancement of the dipole field, and the latter modulates the lower dynamo and the 
amplitude of variations of the quadrupole magnetic field. It can be suggested that the 
neutrino flux from the solar interior is modulated simultaneously. In this case, the phase 
shift of about 1 year between the quadrupole and the neutrino flux variations, on the one 
hand, and the dipole field, on the other, is due to the downward pumping of the dipole 
field, that takes several times as long as the emergence of the quadrupole magnetic tubes 
to the photosphere. The emergence velocity depends on the magnetic field, $B$ 
(according to Parker (1975), $V_b\approx B/4\pi\rho$, where $\rho$ is the matter 
density). Hence, one should expect in the modulation region the existence of strong 
magnetic fields with $T\approx11$ years (probably 10-100 kG) and even stronger 
guadrupole fields, that could ensure a rapid ($le1-2$ months) emergence of the magnetic 
tubes. For the present, it is difficult to estimate the radial dimension of the modulation 
region and the lower dynamo region of the quadruple magnetic field. To judge from the 
relative difference between the real and the model seismic velocities (see Fig. 12 in 
Kosovichev et al., 1997), it makes $\le R_{\odot}$, where $R_{\odot}$ is the radius of 
the Sun.

	The suggested model of generation of the 11-year magnetic cycle by two different 
mechanisms (Rivin, 1998b,c) accounts for the absence of semi-annual wave in the 
$P_{\nu}$ variation, that follows from the VVO hypothesis. This problem is discussed in 
detail in (Bahcall, 1989; Obridko \& Rivin, 1995a, 1996a, 1998c).

	According to VVO model, the neutrino flux is modulated by toroidal components 
of the axisymmetric quasi-dipole magnetic field in the solar convection zone. Only in this 
case, the components of opposite sign in different hemispheres could form an equatorial 
gap, where the field is weak, i.e. the flux modulation must be absent. However the model 
by Rivin (1998b,c) shows that no such gap is possible in the Sun, first, because the field 
in the modulation region is asymmetric between the hemispheres and, second, because 
the equatorial zone is filled with the magnetic tubes of the toroidal quadrupole large--
scale magnetic field emerging from the transition layer to the photosphere.

	With all these considerations taken into account, the VVO model quite adequately 
describes the cyclic modulation and the annual wave of the solar neutrino flux. The 
mechanism of quasi-biennial oscillations needs further investigation on the basis of up-to-
date measurements with various detectors and the analysis of large-scale and local 
(sunspot) magnetic fields in the Sun.

\medskip
\begin{center}
\bf{III. Conclusion}
\end{center}
\smallskip

	The present-day knowledge brings us close to the problem of description 
and simulation of cyclic variations of magnetic fields in the solar interior, however we 
still cannot estimate the magnetic field in the lower dynamo region. Nevertheless, the 
correlation of cyclic variations of $P_{\nu}$  and $\mid B_q\mid$ shows that the main 
characteristics of the magnetic fields, the modulation region in the solar interior, and the 
magnetic moment of the solar neutrino are such as to suggest the reality of the neutrino 
modulation effect. It is the task for the nearest future to more exactly determine each of 
these values from observations of the large-scale magnetic fields on the solar surface, 
from helioseismologic data, as well as by detecting solar neutrino fluxes with up-to--
date facilities that allow the study of their time variations. One of the urgent tasks is to 
analyze the probability of variation with $T\sim27$ days and its subharmonics, and to 
compare this variation (if discovered) with the corresponding variation of the quadrupole 
solar magnetic field. 
\medskip

	The authors are grateful to E.I.Prutenskaya for technical assistance in preparing 
the manuscript.

	The work was done under the sponsorship of the Russian Foundation for Basic 
Research (grant N 96-02-17054) and the State Astronomy Program (grant N 4-264).

\medskip
\begin{center}
\bf{References}
\end{center}
\smallskip

\hangindent=0.5cm \noindent
Akhmedov, E.Kh.// hep-ph/ 9705451. 21.5.1997. 

\hangindent=0.5cm \noindent
Bahcall, J.N.// Neutrino Astrophysics. Cambridge Univ. Press, 1989.

\hangindent=0.5cm \noindent
Davis, R., Cleveland, B.N., Rowley, J.K.// Intersec. Part. and Nucl. Phys. Conf .Steaboat. 
Springs, 23-30 May 1984. New York. 1984. P.1037. 

\hangindent=0.5cm \noindent
Kosovichev, F.G. et al. // Solar Phys. 1997.V.170. P.43. 

\hangindent=0.5cm \noindent
Massetti, S., Storini, M.// Astrophys. J. 1996. V.472. P.287. 

\hangindent=0.5cm \noindent
Neutrino 96 / Eds. Kari Enqvist et al. Helsinki. World Scientific. 1996. 

\hangindent=0.5cm \noindent
Obridko, V.N., Rivin, Yu.R. // Izv. RAN, Seriya.Fizicheskaya.. 1995a. T.59. N9. S.110 
(1995a, Bull.Russ. Acad. Sci. Phys.. 59. N9). 

\hangindent=0.5cm \noindent
Obridko, V.N., Rivin, Yu.R.// Commissions 27 and 42 of the IAU Information. Bulletin 
on Variable Stars 1995b. P.1. 

\hangindent=0.5cm \noindent
Obridko,V.N., Rivin, Yu.R.// Astron.\& Astrophys. 1996a. V.308. P.951. 

\hangindent=0.5cm \noindent
Obridko, V.N., Rivin, Yu.R. //Problemy Geokosmosa. Book of Abstracts. June 17-
23.1996. S.-Petersburg. 1996b. P.90. 

\hangindent=0.5cm \noindent
Obridko, V.N., Rivin, Yu.R.// Astron. J. 1996c. V.73. N 5. S.812 (1996c, Astron. 
Report.40. N5). 

\hangindent=0.5cm \noindent
Parker, E.N. // Astrophys. J. 1975. V.198. P.205. 

\hangindent=0.5cm \noindent
Rivin, Yu.R., Gavryuseva, E.A., Gavryusev, V.G., Kosheleva, L.V. //Geomagnitnye 
variatsij, elektricheskie polja i toki/ Ed. Levitin A.E. M.: IZMIRAN. 1983a. P.153. 

\hangindent=0.5cm \noindent
Rivin Yu.R., Gavryuseva E.A., Gavryusev V.G., Kosheleva L.V. // Issledovanie muonov 
i neitrino v bolshikh vodnykh ob'emakh/ Ed. Kolomeec, Alma-Ata: KazGU. 1983b. 
P.33. 

\hangindent=0.5cm \noindent
Rivin, Yu.R. // Astron. Tsirk. 1989a. N1539. P.22. ( 1989a, Astron. Tsirk. N1539. P.22). 
Rivin, Yu.R. // 5th Simpozium KAPG po solnechno-zemnoi fisike. Samarkand. 1989b. 
P.22. 

\hangindent=0.5cm \noindent
Rivin, Yu.R. // Astron. Tsirk. 1991. N1551. P.26. (1991, Astron. Tsirk. N1551. P.26). 

\hangindent=0.5cm \noindent
Rivin, Yu.R. // Astronom. J.. 1993. V.70. N2. P.392 (1993, Astron. Report. 33. N2). 

\hangindent=0.5cm \noindent
Rivin, Yu.R. // Magnitnye polya Solntaa i gelioseismologiya/ Ed. Dergachev V.A. S-
Petersburg. 1994. p.52. 

\hangindent=0.5cm \noindent
Rivin, Yu.R. // Sovremennye problemy solnechnoiy aktivnosti. S-Petersburg: GAO. 
1997a. Tezisy dokladov. P.76. 

\hangindent=0.5cm \noindent
Rivin, Yu.R. // Sovremennye problemy solnechnoiy aktivnosti /. S-Petersburg.: GAO. 
1997b. Tezisy dokladov. p.80. 

\hangindent=0.5cm \noindent
Rivin, Yu.R. // Sovremennye problemy solnechnoiy aktivnosti / Eds. Makarov V.I., 

\hangindent=0.5cm \noindent
Obridko V.N. S-Petersburg.: GAO. 1997c. P.218. 

\hangindent=0.5cm \noindent
Rivin, Yu.R., Obridko, V.N. // Astronom. J. 1997. V.74. N1. P.83 (1997, Astron. Report. 
41. N1). 

\hangindent=0.5cm \noindent
Rivin, Yu.R. // Izv. RAN, Seriya. Fizicheskaya. 1998a. V.62. N6. P.1263 ( 1998a, Bull. 
Russ. Acad. Sci. Phys. 74. N6)

\hangindent=0.5cm \noindent
Rivin, Yu.R. // Izv. RAN, Seriya Fizicheskaya. 1998a. V.62. N9. P. 1867 (1998b, Bull. 
Russ. Acad. Sci. Phys. 74. N9). 

\hangindent=0.5cm \noindent
Rivin, Yu.R. // Solar Phys. 1998c. (in press) 

\hangindent=0.5cm \noindent
Rivin, Yu.R., Obridko, V.N. // Astronom. J. 1998d. (in press). (1998, Astron. Report. in 
press)

\hangindent=0.5cm \noindent
Sacurai, K. // Nature. 1979. V.278. N5700. P.146. 

\hangindent=0.5cm \noindent
Sacurai, K. // 19th Int. Cosmic Ray Conf., La-Jolla, Aug.11-23, 1985. Washington. 
D.C.1985.P.430. 

\hangindent=0.5cm \noindent
Steshenko, N.V.// Izv. Kryumskoy Astronom. Observatorii. 1967. V.37. P.21. 

\hangindent=0.5cm \noindent
Suzuki, Y.// 17th Int. Conf. On Neutrino Phys. and .Astrophysics. Neutrino-96. Helsinki, 
June 13-19. 1996. World Scientific. 1996. P.73. 

\hangindent=0.5cm \noindent
Voloshin, M.B., Visotsky, M.I., Okun, L.B.// Zh. Teoret. i Experiment. Fiziki, 1986, 
V.91. P.754. 

\hangindent=0.5cm \noindent
Vitinsky, Yu.I., Kopecky, M., Kuklin, G.V. Statistika pyatnoobrazovatelnoi deyatelnosti 
Solntsa M.: Nauka. 1986. 296p. 

\end{document}